\documentclass[twocolumn,showpacs,preprintnumbers,amsmath,amssymb]{revtex4}
\usepackage{graphicx}
\usepackage{dcolumn}
\usepackage{bm}

\begin{document}
\title{Fluctuations in Mass-Action Equilibrium of Protein Binding Networks}
\author{Koon-Kiu Yan$^{1,2}$, Dylan Walker$^{1,2}$, Sergei Maslov$^{2}$}
\affiliation{
$^1$Department of Physics and Astronomy, Stony Brook University, \\
Stony Brook, New York, 11794, USA\\
$^2$Department of Condensed Matter Physics and Material Science, Brookhaven National Laboratory,
Upton, New York 11973,  USA}

\date{\today}
\begin{abstract}
We consider two types of fluctuations in the mass-action equilibrium
in protein binding networks. The first type is driven
by relatively slow changes in total concentrations (copy numbers) of
interacting proteins. The second type, to which we
refer to as spontaneous, is caused by quickly decaying thermodynamic deviations
away from the equilibrium of the system.
As such they are amenable to
methods of equilibrium statistical mechanics
used in our study. We investigate the
effects of network connectivity on these
fluctuations and compare them to their upper and lower bounds. The
collective effects are shown to sometimes lead to large power-law distributed
amplification of spontaneous fluctuations
as compared to the expectation for isolated dimers.
As a consequence of this, the strength of
both types of fluctuations is positively correlated with
the overall network connectivity of proteins forming the complex.
On the other hand,
the relative amplitude of fluctuations is negatively correlated
with the abundance of the complex. Our general findings are
illustrated using a real network of protein-protein interactions
in baker's yeast with experimentally determined
protein concentrations.
\end{abstract}

\maketitle

The study of dynamical fluctuations in complex systems has emerged
as a topic of intense interest germane to the fields of
biology \cite{Bio_fluct}, financial systems \cite{Finance_fluct},
traffic in information \cite{Traffic_info_fluct} and
transportation \cite{Traffic_trans_fluct} networks, and many
others. Of particular interest is the nature of collective effects
that arise as a consequence of the connectivity of the underlying
network. By examining such fluctuations we can understand when the
underlying network plays an important role and when, if possible,
it may be ignored.  A good candidate arena to study dynamical
fluctuations is that of biomolecular processes taking place in cells.

Recently, propagation of biological fluctuations has been studied
in the context of genetic regulation \cite{Noise_gene_net} and
metabolic pathways \cite{Noise_metabolic}.  These studies are
primarily focused on small linear cascades of irreversible
interactions. Conversely, we study the related problem of
fluctuations in the mass-action equilibrium state of
densely-connected, reversible protein-protein-interaction (PPI)
networks.  These networks, in which proteins (nodes) are connected
by edges if they bind together, exhibit non-trivial topological properties such
as clustering and loops of
various lengths.  Ourselves and others have studied the effect of
large systematic changes in the levels of just one or a few proteins
on the mass-action equilibrium
of their PPI networks \cite{Sergei_NJP,Sergei_PNAS}. Such changes are
likely to occur as a consequence of regulated response of the cell to
big changes in the external environment.
For the same system, however, there is another type of
perturbation that is both different and of significant interest:
intracellular noise or small fluctuations in equilibrium (bound and free)
concentrations of many proteins.
The randomness, smaller magnitude, and sheer number of involved proteins characterize
the difference between the noise and fluctuations that are the subject of this study and the systematic
large changes in the total abundance single proteins that we recently studied in
\cite{Sergei_PNAS}. These fluctuations come in two varieties.
\textit{Spontaneous} fluctuations in bound concentration occur at
constant protein copy number, due to the intrinsic
stochastic nature of binding interactions. These fluctuations are small but
change rapidly relative to the characteristic time of changes in
protein copy number.
They are well described using the machinery of equilibrium statistical physics.  In
contrast, \textit{driven} fluctuations are induced by changes in
protein copy number due to the stochastic
nature of their production and degradation as well as
variation in activity of global factors controlling the overall protein abundance.
This driven noise is usually larger than the
spontaneous noise. It also happens on timescales (tens of minutes) that are large
compared to the relaxation time of the mass action equilibrium
which are rarely slower than seconds.
In fact, the problem of thermal
noise is well connected to the static response of the system to
systematic concentration changes, as we will show.

To illustrate general principles with a concrete example, in this study,
we used a highly curated genome-wide network of PPI in yeast ({\it S.
cerevisiae}), which, according to the BIOGRID database \cite{BIOGRID}, were
independently confirmed in at least two publications. We combined
this network with a genome-wide data set of protein abundances
\cite{ghaemmaghami2003gap}. After keeping only the
interactions between proteins with known concentrations, we were
left with 4,185 binding interactions among 1,740
proteins. The same network was previously used by us
and others in \cite{Sergei_NJP,Sergei_PNAS}.
Another simplification (partially justified in these earlier studies) is that:
1) we consider only homodimers and heterodimers and thus ignore the formation of higher order complexes, 2) we use the same dissociation constant $K_{ij}=1$nM (or 34 proteins/cell) for all interactions in our network. This choice is justified by a relative lack
of sensitivity of equilibrium concentrations to details of assignment of dissociation constants to individual interactions (see Fig. 5 in Ref. \cite{Sergei_PNAS}).

The PPI network defines the backbone of a system of dynamical
dimer formation and dissociation.  The system at any instant is
described by $\{C_i\}$, the set of total protein concentrations,
$\{D_{ij}\}$, the concentrations of all dimers $(ij)$ and
$\{F_i\}$, the set of free protein concentrations.  We will assume
reactions are occurring in a unit volume, in order to suppress the
system volume $V$ in the equations that follow.  At any instant,
the system is constrained by mass conservation:
\begin{equation}
\label{mass_conservation}
C_i = F_i + \sum_j{D_{ij}}
\end{equation}
so that $F_i$ is not an independent variable.  For considerations
of noise we use deviations of dimer concentration, $\delta
D_{ij}$, away from their long term averages.  The second moments of these fluctuations, $\langle\delta D_{ij}^2\rangle$ quantify the strength of the noise.

\begin{figure}[!htb]
\includegraphics*[width=\linewidth]{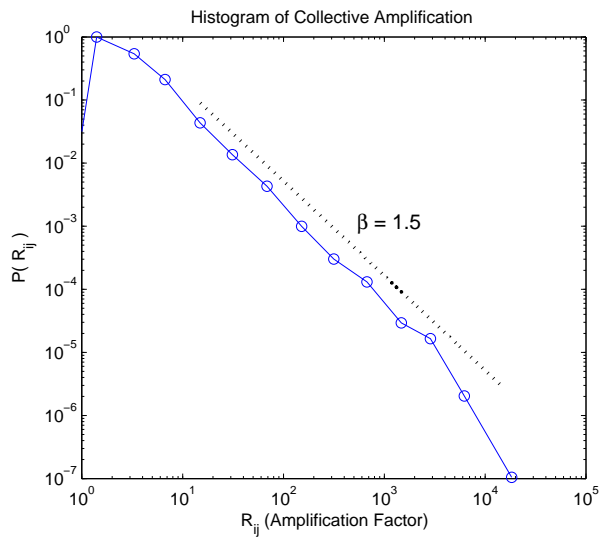}
\caption{\label{fig1}
Histogram of amplification factors for spontaneous (thermal) noise
of equilibrium dimer concentrations $D_{ij}$ in the
PPI network of yeast.  Collective effects lead to amplification
relative to the isolated dimer null model.  Large
amplification, up to several orders of magnitude,
is sometimes achieved.}
\end{figure}

To study \textit{spontaneous} fluctuations, we consider the case
where all total concentrations $C_i$ are held constant and variations in free
and dimer concentration are driven solely by thermal fluctuations.  To
this end, we write the partition function for a network
of interacting dimers:
\begin{equation}
Z=\exp(-G/k_BT)
=\sum_{\{D_{ij}\}}{{N_S(\{D_{ij}\})\exp(-\sum_{i<j}{\frac{\epsilon_{ij}
D_{ij}}{k_BT}})}}
\label{statsum}
\end{equation}
where the sum is taken over all possible (integer) copy numbers of individual dimers defining
the  ``occupational state''  $\{D_{ij}\}$. The
combinatorial factor $N_S(\{D_{ij}\})$ counts the number of microstates
of individual labeled proteins resulting in a given
occupation state $\{D_{ij}\}$. For example, for a
single dimer $AB$, $N_S(D_{AB})$ is the combinatorial factor:
\begin{equation}
N_S(D_{AB})={C_A \choose D_{AB}}{C_B \choose D_{AB}}D_{AB}!=\frac{C_A!C_B!}{D_{AB}!F_A!F_B!}
\end{equation}
Using the Stirling's approximation for factorials in
$N_S(\{D_{ij}\})$ one gets a concise expression for the
free energy for an arbitrary network of dimers:
\begin{eqnarray}
\nonumber   G&=& \sum_{(ij)\epsilon E}\{\epsilon_{ij} D_{ij} + k_BTD_{ij}[\log(D_{ij})-1]\}\\
\nonumber    &+&k_BT\sum_{i=1}^N\{F_i[\log(F_i)-1]-C_i[\log(C_i)-1]\}\\
\end{eqnarray}
where the first sum runs over all $E$ edges (dimers) and the
second sum runs over all nodes (proteins) in the network.
Free concentrations $F_i$ in this expression are not
independent variables but rather a shorthand for
$C_i-\sum_m D_{im}$.
The above expression does not include volume-dependent entropy and
kinetic terms that we have suppressed as they are not relevant to
our discussion here.
The first derivative of the free energy with respect to
dimer concentration gives the Law of Mass Action (LMA)
that relates free and bound equilibrium concentrations in the system via
$D_{ij}=F_iF_j/K_{ij}$, where $K_{ij}=K^{(0)}\exp (-\epsilon_{ij}/k_BT)$.
The second derivative of the free energy with respect to dimer
concentration yields the generalized susceptibility and, in
accordance with the Fluctuation-Dissipation Theorem (FDT), the
noise
\begin{equation}
\eta\equiv \frac{\langle \delta D_{ij}^2 \rangle}{D_{ij}}=\left(\Gamma^{-1}\right)_{(ij)(ij)}
\end{equation}
where
\begin{eqnarray}
\nonumber \Gamma_{(ij)(km)} &=& \frac{D_{ij}}{k_BT} \frac{\partial^2 G}{\partial D_{ij} D_{mk}}\\
\nonumber                   &=& \delta_{ik}D_{ij}/F_i + \delta_{jm}D_{ij}/F_j + \delta_{ik}\delta_{jm}.
\end{eqnarray}
A direct consequence of this result is that spontaneous
fluctuations for a dimer linked to the rest of the network
involve contributions from
other dimers, through the inverse of $\Gamma$, the so-called
collective effects of the network.  To address the impact of
collective effects on the noise, it seems natural to compare the
noise of a dimer in the network to the noise for an isolated dimer
(\textit{isol-F}) with the same equilibrium concentrations $F_i$, $F_j$, and
$D_{ij}$. Such an isolated dimer corresponds to a matrix $\Gamma$
that is diagonal and has a trivial inverse such that:
\begin{equation}
\eta^{\mathrm{isol-F}}=[\Gamma_{(ij)(ij)}]^{-1}=[\frac{D_{ij}}{F_i} + \frac{D_{ij}}{F_j} + 1]^{-1}
\end{equation}
Furthermore, $\eta>\eta^{\mathrm{isol-F}}$ is easily shown from
the convexity of $\Gamma$ \cite{lower_limit}.  Clearly then,
collective effects act to amplify thermal fluctuations.  This is
related to propagation of static perturbations, studied in
\cite{Sergei_NJP}, as fluctuations from neighboring dimers
contribute to a dimer's own noise.  We define the amplification
factor for a dimer $(ij)$:
\begin{equation}
R=\eta/\eta^{\mathrm{isol-F}}
\end{equation}
A histogram of amplification factors for the PPI network of
baker's yeast is examined in Fig. \ref{fig1}.
Relative to the isolated case, collective amplification can lead
to thermal noise that is orders of magnitude larger, as is evident
from this histogram.
The distribution has a
power-law tail with an approximate exponent of $\beta \simeq
1.5$.

\begin{figure}[!htb]
\includegraphics*[width=\linewidth]{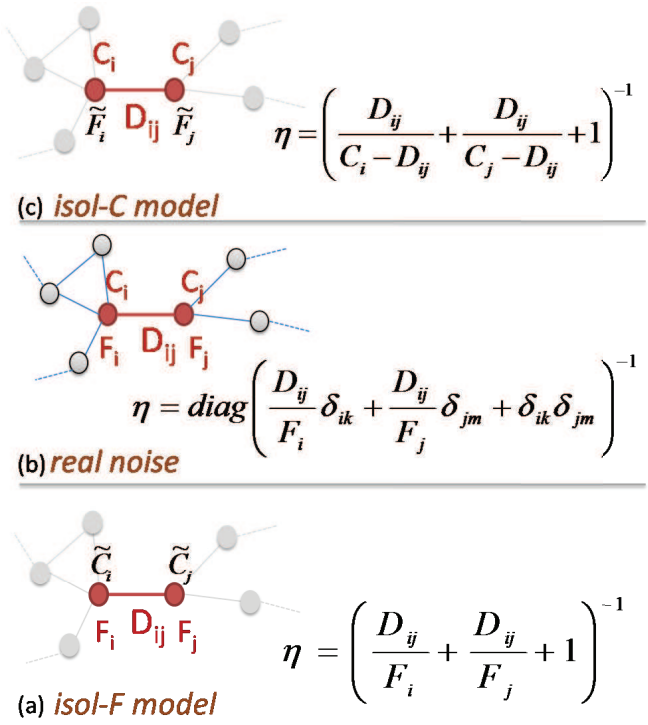}
 \caption{\label{fig2} A comparison of the noise in a network
 dimer to two isolated dimer models defined in the text. (a) The
 \textit{isol-F} model: Each dimer $(ij)$ is isolated and has the
 same protein free- ($F_i$, $F_j$) and dimer- concentrations
 ($D_{ij}$) as the corresponding dimer in the network.  This model
 ignores the contribution of other dimers to the noise of dimer
 $(ij)$
 (b) The noise of a dimer $(ij)$ in the network is given by
 the $(ij),(ij)$ diagonal element of the inverse of the matrix
 $\Gamma$ as described in the text.
 (c) The \textit{isol-C} model: Each dimer $(ij)$ is isolated and
 has the same protein total- ($C_i$, $C_j$) and dimer-
 concentrations ($D_{ij}$) as the corresponding dimer in the
 network.  The real noise is bound below and above by the isolated models $\eta^{\mathrm{isol-F}}<\eta<\eta^{\mathrm{isol-C}}$}
\end{figure}

\begin{figure}[!htb]
\includegraphics*[width=\linewidth]{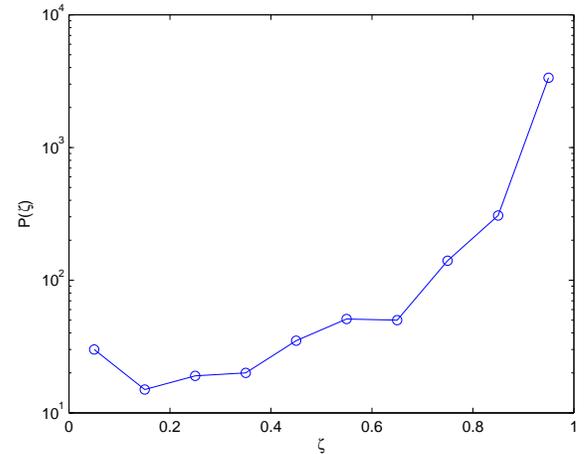}
\caption{\label{fig3}
Histogram of the spontaneous noise coordinate $\zeta$ in
the PPI network of yeast.  The coordinate describes the position of
noise amplitude relative to its lower ($\zeta=0$) and
upper ($\zeta=1$) limits described in the text.
}
\end{figure}

Collective amplification of thermal noise presents a worrisome
theoretical possibility.  Can amplification occur without limit?
To address this question, it is fruitful to develop an alternative
formalism in which the magnitude of fluctuations are calculated
directly from the partition function.
Using Eq. \ref{statsum} it
is straightforward to show, by a change of variables, that
higher moments of $D_{ij}$ can be related to the lower moments
evaluated at a reduced system size.  In particular:
\begin{equation}
\nonumber \langle D_{ij}(D_{ij}-1)\rangle = \langle D_{ij} \rangle \langle D_{ij} \rangle |_{C_i-1, C_j-1}
\end{equation}
where the latter moment is evaluated in a system for which the
copy number of proteins $i$ and $j$ ($C_i$ and $C_j$)
are reduced by exactly one.
It follows that the noise may be alternatively expressed as:
\begin{equation}
\eta = 1 + \langle D_{ij}\rangle |_{C_i-1, C_j-1} - \langle D_{ij} \rangle
\end{equation}
 The above expression for thermal noise hints at a connection to
 static perturbations of total concentration.  This connection can
 be made even more explicit by expanding the 2nd term to first
 order in total concentration:
\begin{equation}
\label{eta_taylor_expand}
\eta \simeq 1 - D_{ij}[(\Lambda^{-1})_{ii} + (\Lambda^{-1})_{jj} +2(\Lambda^{-1})_{ij}]
\end{equation}
where the matrix:
\begin{equation}
\nonumber\Lambda_{ij} = \frac{\partial C_i}{\partial \log F_j}= D_{ij}+C_i\delta_{ij}
\end{equation}
characterizes the response of equilibrium concentrations $F_m$ to small static changes
in total concentrations $C_k$ \cite{Sergei_NJP}.
It should be remarked that, despite the approximation used in Eq.
\ref{eta_taylor_expand}, this approach is in good agreement with
the FDT formalism first introduced. One notes that this expression for
noise explicitly depends only on the total and dimer concentrations
used to define the matrix $\Lambda$.  This suggests the definition
of a new isolated model (\textit{isol-C}), consisting of an isolated $(ij)$
dimer formed by proteins with the same
$C_i$, $C_j$ and $D_{ij}$.
This is only possible
through changes in the dissociation constant and free
concentrations of constituent proteins $i$ and $j$.
It is important to mention that this
model is distinct from the \textit{isol-F}
benchmark defined earlier, in which each
isolated dimer has the same equilibrium free and dimer
concentrations (yet different $C_i$ and $C_j$)
as the corresponding dimer in the network.  For an \textit{isol-C}
dimer, the matrix $\Lambda$ is 2x2 and trivially
invertible.  The noise is given by:
\begin{equation}
\eta^{\mathrm{isol-C}}=\left(\frac{D_{ij}}{C_i - D_{ij}} + \frac{D_{ij}}{C_j - D_{ij}} + 1 \right)^{-1}
\end{equation}
A comparison with the \textit{isol-F} model reveals that a dimer
in the \textit{isol-C} model has an equilibrium free concentration
$\tilde{F_i} = F_i + \sum_{k}{D_{ik}}$ and similarly for protein
$j$. In other words, the contribution of neighboring dimers to the
noise of dimer $(ij)$ has been included by absorbing them into an
effective free concentration.
Moreover, where the \textit{isol-F} model completely ignores the
effect of neighboring dimers, the \textit{isol-C} model brings
neighboring sources of noise one step closer to dimer $(ij)$.
Consequently, the noise of a dimer in the \textit{isol-C} model
always exceeds the noise of a corresponding dimer in the real
network.  The real noise for a dimer in a network falls between
the bounds of the two isolated dimer noise predictions.  A
summary of the lower- and upper-bound models and their noise is
given in Fig. \ref{fig2}.
The actual spontaneous fluctuations achieved
are a result of real network topology and the distribution of
total protein concentration.  It is natural to ask how these
fluctuations compare to their minimally and maximally achievable
values.  This suggests the coordinate transformation:
\begin{equation}
\nonumber\eta \equiv (1-\zeta)\eta^{\mathrm{isol-F}} + \zeta \eta^{\mathrm{isol-C}}
\end{equation}
A histogram of $\zeta$ for the PPI network of yeast is shown in Fig. \ref{fig3}.
Of particular note is the large pileup against the
upper limit of amplification.  In real PPI networks,
it would seem that collective effects lead to
amplification quite close the maximally achievable
limit.

Now we turn our attention to the second type of noise
\textit{driven} by stochastic changes in
protein copy number $C_i$. In the living cell,
these fluctuations are typically much larger than the
spontaneous fluctuations. Furthermore, the changes in $C_i$
occur at a relatively slow time scale (tens of minutes)
so that the mass-action equilibrium respond to these changes.
From the results of \cite{Sergei_NJP,Sergei_PNAS} it follows that,
in general, the amplitude of driven fluctuations is given by:
\begin{equation}
\frac{\langle\delta D_{ij}^2 \rangle}{D_{ij}^2}=\langle ( \sum_k{(\Lambda^{-1})_{ik}} \delta C_k +\sum_m{(\Lambda^{-1})_{jm} \delta C_m} )^2 \rangle
\end{equation}
The evaluation of the above expression requires
the full matrix of cross-correlations $\langle \delta C_k \delta C_m
\rangle$ which is currently experimentally unknown.
For the simplest
case $\langle \delta C_k \delta C_m \rangle \propto
C_k^2 \delta_{mk}$ of uncorrelated fluctuations
(so-called intrinsic noise
\cite{Bio_fluct}), the driven response becomes:
\begin{equation}
\left(\frac{\langle\delta D_{ij}^2 \rangle}{D_{ij}^2}\right)_{\mathrm{int}}\propto \sum_k{[(\Lambda^{-1})_{ik} + (\Lambda^{-1})_{jk}]^2C_k^2}
\end{equation}

In conclusion we study how the two types of
noise studied above relate to simple predictors such as
abundance and connectivity (number of connections a
dimer has to the network).  With high statistical
significance, we find that the relative amplitude
($\sqrt{\langle\delta D_{ij}^2 \rangle}/D_{ij}$)
of both spontaneous and driven (intrinsic) noise
is negatively correlated with dimer abundance $D_{ij}$
(Spearman coefficient of $r=-0.98$, $r =-0.64$, respectively).
This result is generally expected for relative noise
amplitudes due to the law of large numbers. Indeed, for
independent fluctuations it is expected to decrease as
$\sim 1/\sqrt{D_{ij}}$. This explains a particularly
strong correlation in the case of spontaneous
fluctuations.  Furthermore, we found that relative
amplitude of  both spontaneous and driven (intrinsic) noise are positively correlated with
connectivity ($r=0.46$, $r=0.34$). This is consistent with the
overall scenario that we investigated above in which
any type of noise propagates throughout the network and network connections
(both direct and, to some extent, indirect) to noisy partners
positively contribute to fluctuations of individual dimers.

Work at Brookhaven National Laboratory was carried out
under Contract No. DE-AC02-98CH10886, Division of
Material Science, U.S. Department of Energy.



\begin{thebibliography}{3}
\bibitem{lower_limit} For brevity, assume the notation $\mu=(ij)$, $\nu=(mk)$.  The matrix $\Gamma$ is symmetrized by the diagonal matrix $Q_{\mu\nu} = \sqrt{D_{\mu\nu}}\delta_{\mu\nu}$ and is diagonalized by a unitary transformation $U$ so that: $\Gamma = QU\Gamma_DU^{-1}Q{-1}$.  From these considerations and the convexity of $f(x)=x^{-1}$ it can be shown that $\Gamma^{-1}_{\mu\mu}\geq (\Gamma_{\mu\mu})^{-1}$,  where $\Gamma^{-1}_{\mu\nu} = \sum_{\alpha}{U^2_{\mu \alpha}(\Gamma_D)^{-1}_{\alpha \alpha}}$ and $(\Gamma_{\mu\mu})^{-1}=(\sum_{\alpha}{U^2_{\mu \alpha} (\Gamma_D)_{\alpha \alpha}})^{-1}$.
\end{thebibliography}
\end{document}